\documentclass{osa-article}

\journal{oe}

\articletype{Research Article}

\usepackage{lineno}

\usepackage{xcolor}
\usepackage{graphicx}
\usepackage{dcolumn}
\usepackage{bm}
\usepackage{setspace}
\usepackage{hyperref}
\usepackage{url}
\usepackage{siunitx}
\usepackage[title]{appendix}
\usepackage{algpseudocode}
\usepackage{algorithm}

\newcommand{\bk}{\mathbf{k}}
\newcommand{\br}{\mathbf{r}}

\newcommand{\fft}[1]{\mathcal{F} \left\{ #1 \right\}}
\newcommand{\ifft}[1]{\mathcal{F}^{-1} \left\{ #1 \right\}}

\newcommand{\expp}[1]{\exp \left( #1 \right)}
\newcommand{\figref}[1]{Fig.~\ref{#1}}

\newcommand{\secref}[1]{Sec.~\ref{#1}}

\renewcommand{\eqref}[1]{Eqn.~\ref{#1}}

\begin{document}

\title{Addressing phase-curvature in Fourier ptychography}
\author{Tomas Aidukas\authormark{1,2,*}, Lars Loetgering\authormark{3,4,5}, Andrew R. Harvey\authormark{1}}

\address{\authormark{1}School of Physics and Astronomy, University of Glasgow, G12 0QQ, Glasgow, UK\\
\authormark{2}Paul Scherrer Institute, PSI CH, Forschungsstrasse 111, 5232 Villigen, Switzerland\\
\authormark{3}Leibniz Institute of Photonic Technology, Albert-Einstein-Straße 9, 07745, Jena, Germany\\
\authormark{4}Helmholtz-Institute Jena, Fröbelstieg 3, 07743 Jena, Germany\\
\authormark{5}Institute of Applied Physics and Abbe Center of Photonics, Friedrich Schiller University Jena, Albert-Einstein-Straße 15, 07745 Jena, Germany}
\email{\authormark{*}tomas.aidukas@psi.ch} 

\begin{abstract}
In Fourier ptychography, multiple low resolution images are captured and subsequently combined computationally into a high-resolution, large-field of view micrograph. A theoretical image-formation model based on the assumption of plane-wave illumination from various directions is commonly used, to stitch together the captured information into a high synthetic aperture. The underlying far-field (Fraunhofer) diffraction assumption connects the source, sample, and pupil planes by Fourier transforms. While computationally simple, this assumption neglects phase-curvature due to non-planar illumination from point sources as well as phase-curvature from finite-conjugate microscopes (e.g., using a single-lens for image-formation). We describe a simple, efficient, and accurate extension of Fourier ptychography by embedding the effect of phase-curvature into the underlying forward model. With the improved forward model proposed here, quantitative phase reconstruction is possible even for wide fields-of-views and without the need of image segmentation. Lastly, the proposed method is computationally efficient, requiring only two multiplications: prior and following the reconstruction. 
\end{abstract}

\section{Introduction}

Fourier ptychographic microscopy (FPM) is a computational-imaging technique developed for wide-field, high-resolution phase-contrast imaging~\cite{Zheng2013Wide, Zheng2021Concept, Konda2020Fourier}, to overcome the bandwidth limitation imposed by the finite aperture of a microscope. While in traditional microscopy a higher numerical aperture (NA) enables a higher spatial-frequency cut-off (and thereby higher-resolution images), it comes at the cost of reduced field-of-view (FoV). The inverse relationship between FoV and resolution is imposed by the space-bandwidth-product (SBP), which defines the total number of independent pixels that can be captured by an imaging system~\cite{Konda2020Fourier}. While band-pass filtering is unavoidable, FPM overcomes the frequency cut-off by capturing a sequence of low-resolution bright-field and dark-field images using variable illumination angles. Under angular illumination, each of these images represent partially overlapping frequency bands of the wideband sample spectrum. During image reconstruction, these diverse images are  coherently combined in the Fourier-domain into a wideband image spectrum.  The process of synthesising a high-resolution image requires determination of the phases of the angularly-captured images, which are lost during image detection. Phase retrieval thus forms the core of FPM reconstruction algorithms. Image reconstruction can be considered as an optimization problem, in which the difference is minimized between the experimental measurements and the expected intensity given by the theoretical image-formation model. Even for ideal noiseless data, the reconstructed image quality can be compromised if the underlying forward model is inaccurate. In this manuscript, we address two likely sources of error within the FPM image-formation model, which can lead to severe degradations of the reconstructed images if not addressed properly.


Firstly, the commonly used forward model in FPM describes the sample and pupil plane as Fourier conjugates~\cite{Zheng2013Wide}, but this is strictly true only for telecentric imaging system~\cite{Mertz2019Introduction}. Telecentricity provides constant magnification across the field-of-view, which is typically achieved with microscope objectives within high-quality infinite-conjugate microscopes. However, the use of well corrected objectives can be impractical due to their price for low-cost applications~\cite{Dong2014FPscope,Aidukas2019Low,Chan2019Parallel} or other unconventional configurations such as multi-lens, multi-camera microscopy~\cite{Aidukas2021Next,Aidukas2021High,Aidukas2019Multi}.  We will show that a non-telecentric imaging system (e.g., using a single-lens for image-formation) contains additional phase-curvature terms in the forward image-formation model. While telecentricity has been used to eliminate phase-curvature in digital holographic microscopy~\cite{Doblas2015Physical,Ferraro2003Compensation}, the distinction between optical configurations and their corresponding forward models has not been made in the context of FPM.

Secondly, the conventional FPM model assumes ideal plane-wave illumination~\cite{Zheng2013Wide, Zheng2013Characterization, Ou2013Quantitative, Yeh2015Experimental, Song2019Full,Konda2020Fourier,Zhu2020Single, Zheng2021Concept}. This assumption applies to certain FPM implementations where collimated laser or synchrotron radiation was used~\cite{Wakonig2019X,Kuang2015Digital}. However, to date the most common illumination source found in FPM systems are LEDs~\cite{Tian2014Multiplexed,Dong2014FPscope,Tian2015Computational,Phillips2015Multi,Horstmeyer2015Digital,Guo2016Fouriera,Phillips2017Quasi,Eckert2018Efficient,Aidukas2019Low,Konda2020Fourier,Aidukas2021Next,Zheng2021Concept}, which are more appropriately modelled by spherical-wavefronts~\cite{Goodman2005Introduction}. To approximate a point-source illumination by a plane-wave, the source should be positioned sufficiently far away from the sample such that the curvature of the illumination wave front is negligible over the field of view of interest~\cite{Zheng2013Wide,Yeh2015Experimental}. This is not possible for compact experimental setups~\cite{Dong2014FPscope,Aidukas2019Low}, unless infinite-conjugate microscopes are used. In addition, a very large source-to-sample distance would compromise the illumination NA (assuming source hardware of fixed size), leading to reduced resolution of the reconstructed images. 

The two phase-curvature contributions mentioned above can be partially mitigated computationally, by partitioning the image field of view into smaller segments~\cite{Zheng2013Wide, Zheng2013Characterization,Song2019Full,Ou2014Embedded, Tian2015Computational, Tian2014Multiplexed}, over which the plane-wave approximation is more accurate~\cite{Zheng2013Wide}. When each of the smaller field of views is reconstructed, the results are then stitched together into a single wide-field, high-resolution image.  While  segmentation-based reconstruction was introduced to address non-planar wavefronts, all field-of-view dependent phenomena are reduced, including the effects of non-telecentric optics. This reconstruction approach also alleviates the issue of space-variant point-spread functions because aberrations can be retrieved independently for each segment~\cite{Zheng2013Characterization,Ou2014Embedded,Song2019Full}. In addition, segmenting the FoV allows for distributed data inversion across multiple processing units~\cite{Nashed2014Parallel,Wakonig2020PtychoShelves}, resulting in an increase of computational speed. However, even with segmentation-based reconstruction, the exclusion of the phase-curvature terms from the forward model can lead to poor algorithmic convergence, phase inconsistencies between adjacent segments and overall phase aberrations, all of which will be shown in the following sections.

By incorporating the aforementioned quadratic phase contributions to the FPM forward model, reconstruction quality and convergence speed can be improved. We provide both simulations and experimental results indicating that although artefacts due to neglected phase-curvature in the forward model are largely mitigated in segmentation-based FPM, they cannot fully be eliminated. Instead, we  demonstrate that significant phase aberrations can be eliminated using our proposed method when either the plane-wave and/or telecentric imaging assumption is violated. Our proposed computational correction requires only two multiplications prior and after the reconstruction, resulting in a minor increase in computational complexity while drastically improving the quality and reliability of reconstructed images.

We  begin by introducing the wave-optical description of the FPM image-formation model and highlight the phase-curvature terms. Computational reconstruction and corrections will be explained, followed by validation of the phase-curvature correction using simulated and experimental data. Lastly, the findings are discussed and concluded.

\section{Wave-optical FPM model}
\begin{figure}[t]
    \centering
    \includegraphics[width=\linewidth,trim={0 0 0 0},clip]{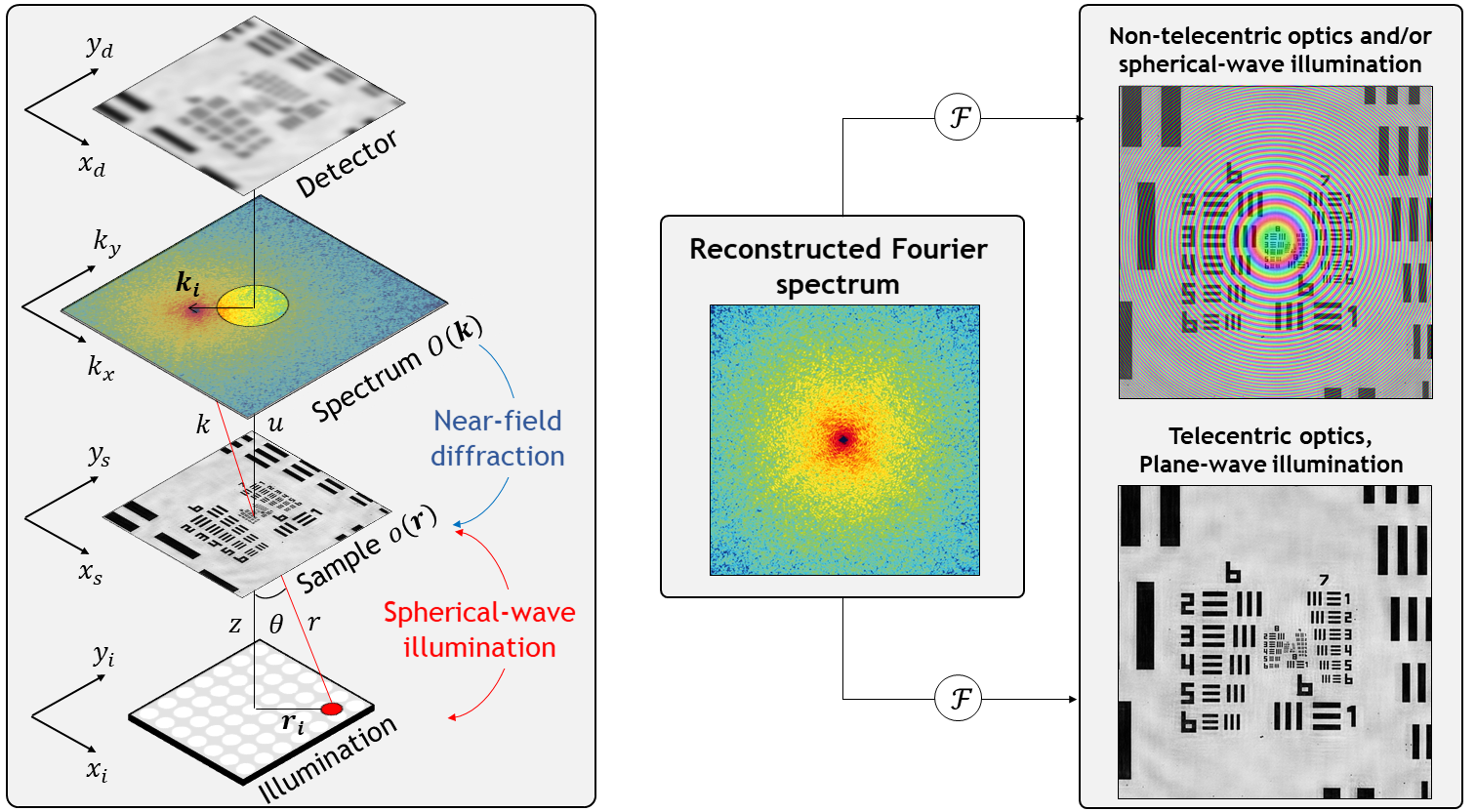}
    \caption{The origin of the phase-curvatures is due to spherical-wave illumination and sample-to-lens wave-propagation in a non-telecentric imaging system. phase-curvature severity will ultimately depend on the propagation distances $u$ and $z$, and the FoV area being imaged. During iterative Fourier spectrum reconstruction, both of these phase-curvatures will be recovered. By propagating the reconstructed spectrum to the sample plane, the phase-curvatures will reveal themselves as undesirable phase aberrations.}
    \label{fig:figure1}
\end{figure}

The standard image-formation model in FPM assumes that the sample $o(\br)$ is illuminated with plane-waves to produce the diffracted sample spectrum $O(\bk)$, illustrated by \figref{fig:figure1}. With plane-wave illumination, the sample spectrum is translated in the pupil plane by $\bk_i$. The translated spectrum is low-pass filtered by the pupil $P(\bk)$ and propagated to the image plane by the Fourier transform $\fft{\cdot}$. The detected intensity with illumination by the $i^{\text{th}}$ LED is~\cite{Zheng2013Wide}: 
\begin{equation}
    \begin{split}
        I_{i}(\br) &=  |\fft{ P(\bk) O(\bk - \bk_{i}) }|^2.
    \end{split}
    \label{eqn:image_formation}
\end{equation}
By synthesizing multiple experimental images in the Fourier-domain (captured with angular illumination), a broadband spectrum $O(\bk)$ can be reconstructed to produce a high-SBP sample image $o(\br)$. Typically, the far-field diffraction assumption is used to describe the transformation between the reconstructed sample image and its spectrum~\cite{Zheng2013Wide,Horstmeyer2016Computational}:
\begin{equation}
    O(\bk) = \fft{o(\br)}.
    \label{eqn:O_farfield}
\end{equation}
To accommodate non-planar illumination wavefront and non-telecentric imaging, we re-derive the wave-propagation and diffraction process starting from the illumination source and ending at the detector. By using Fresnel diffraction for wave-propagation between optical components and assuming spherical (rather than planar) illumination wavefronts we demonstrate in \secref{section:image-formation model} that the FPM forward model in \eqref{eqn:image_formation} is valid, provided that the diffracted spectrum satisfies:
\begin{equation}
    O(\bk) = \fft{o(\br)Q(\br)}.
    \label{eqn:O_nearfield}
\end{equation}
Compared to \eqref{eqn:O_farfield}, the relationship between the sample and its spectrum in \eqref{eqn:O_nearfield} contains a phase-curvature term $Q(\br)$, which will depend on the imaging configuration being used:
\begin{align}
    Q(\br) = \expp{ik\left(\frac{1}{2u} + \frac{1}{2z} \right)(x^2+y^2)}.
    \label{eqn:quadratic_phase}
\end{align}

Here, $z$ is the distance from the LED to the sample plane and $u$ is the distance from the sample to the pupil plane, where we assume a single-lens FPM (non-telecentric) system. In a non-telecentric, single-lens imaging system, the expected phase-curvature will be the result of wave-propagation (proportional to $1/u$), as illustrated in \figref{fig:figure1}. In the presence of spherical illumination, an additional phase-curvature (proportional to $1/z$) will appear. In a telecentric imaging system, the only possible contributions to the observed phase-curvature are due to non-planar illumination wavefronts. Since the illumination-to-sample distance ($z$) will be almost certainly longer than the working distance of a microscope ($u$), lesser phase-curvature is expected in a telecentric imaging system. However, overall phase-curvature is generally significantly greater in compact imaging systems~\cite{Dong2014FPscope,Aidukas2019Low,Chan2019Parallel}, irrespective of the microscope objective being used. 

The phase-curvature can be reduced by manipulating the distances $u$, $z$ and the FoV area $(x^2+y^2)$ inside \eqref{eqn:quadratic_phase}. The sample-to-lens distance $u$ cannot be changed easily, since it defines the desired optical magnification and the working distance of the microscope. The same applies to the illumination propagation distance $z$, which will affect the maximum illumination NA that can be synthesized. The reconstructed FoV area is the only quantity that can be adjusted without modifications of the experimental setup. By segmenting the image FoV into tiles of a desired size, we can reduce the expected phase-curvature.

Lastly, based on \eqref{eqn:image_formation} we see that recording data in the image plane is not directly sensitive to the phase-curvature $Q(\br)$. In particular, the absolute value squared operation renders a direct observation of the quadratic phase impossible. It may thus seem that we could simply ignore it, without affecting the underlying FPM model. However, the quadratic phase within $Q(\br)$ does affect the reconstruction of the spectrum $O(\bk)$, which FPM seeks to stitch together in a self-consistent way. We will demonstrate below that neglecting $Q(\br)$  leads to a poor initialization of the underlying forward model, which can prevent convergence to a feasible solution.

\section{Computational methods}
\subsection{Fourier ptychography reconstruction}
FPM reconstruction can be regarded as a cost-function minimization problem between the experimental observations and parameters being estimated~\cite{Odstrcil2018Iterative,Yeh2015Experimental,Thibault2012Maximum}. One example of such a cost function for a given illumination $i$ can be written as the L2-norm between the measured and expected amplitude:
\begin{equation}
    \mathcal{L}_i = \left|\left| \sqrt{I_{i}(\br)}  - |\fft{ P(\bk) O(\bk - \bk_{i}) }| \right|\right|^2.
    \label{eqn:cost_fn}
\end{equation}
How such optimization should be carried out is outside the scope of this manuscript, and one should instead refer to one of the following texts~\cite{Odstrcil2018Iterative,Yeh2015Experimental,Thibault2012Maximum,Aidukas2021Next}. Provided that the forward model is correct, we can then successfully recover both $O(\bk)$ and $P(\bk)$. However, based on \eqref{eqn:O_nearfield} the reconstructed spectrum will include the quadratic phase factors $Q(\br)$, which will be coupled with the recovered sample function $o(\br)$. To achieve an aberration-free reconstruction, the quadratic phase factors from \eqref{eqn:quadratic_phase} must be eliminated.

\begin{figure}[t]
    \centering
    \includegraphics[width=\linewidth,trim={0 0 0 0}]{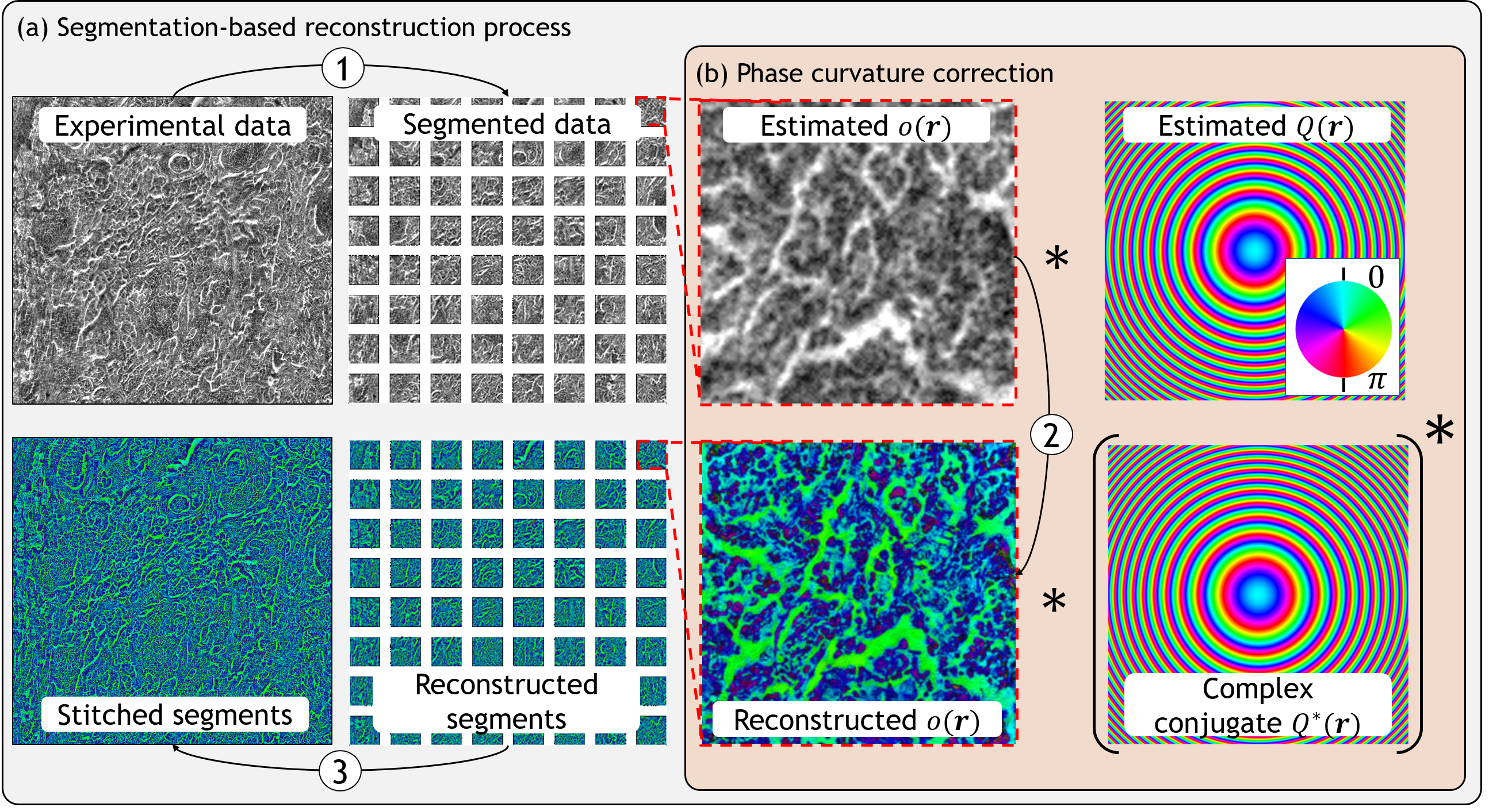}
    \caption{Example of the segmentation-based reconstruction framework in (a). Step 1: divide FOV into small segments. Step 2: reconstruct each FOV segment. Step 3: stitch all reconstructed segments into a single wide-field, high-resolution image. phase-curvature correction in (b) is performed by initializing the reconstruction with the predicted phase-curvature and conjugating it from the reconstruction.}
    \label{fig:fpm_reconstruction}
\end{figure}

As noted previously, a narrow image field-of-view can be used to minimize the quadratic phase exponentials. This can be done by splitting the image FoV into tiles and reconstructing multiple image spectra $O(\bk)$, each corresponding to a fraction of the sample $o(\br)$ being reconstructed~\cite{Zheng2013Wide}. Once all of the segments are reconstructed, they are stitched into a single wide-field, high-resolution image. We will refer to such reconstruction process as “segmentation-based reconstruction” which is illustrated by \figref{fig:fpm_reconstruction}(a). Each segment will also have a unique pupil function $P(\bk)$ to account for field-varying aberrations. In principle, each FOV segment can be regarded as a “mini-experiment” with its own unique aberrations $P(\bk)$, spatial frequency sampling vectors $\bk_i$. However, we will show that even with small segment sizes the classical FPM forward model (\eqref{eqn:image_formation} and \eqref{eqn:O_farfield}) is not guaranteed to yield the desired reconstructions, unless the quadratic phase exponentials $Q(\br)$ are mitigated.

\subsection{Phase-curvature correction}
Whatever the reconstructed segment size may be, some residual or even severe phase-curvature will remain. To eliminate it, we propose a simple computational method based on initialization of the reconstructed object with the theoretical phase-curvature and its removal after the reconstruction is finished. Since the optimization landscape in FPM is non-convex, poor initialisation of the sample and pupil can result in slow convergence or stagnation in a local minimum. To push the algorithm closer towards the global minimum, a good initialization is crucial~\cite{Fienup1982Phase,Tian2015Computational}. We obtain an initial estimate for our FPM reconstruction by means of the following steps. First, the mean over all captured images is computed. Second, the aforementioned mean image is upsampled to the pixel size of the high-NA synthetic Fourier space resulting from all illumination directions, to produce $o_{\mathrm{init}}(\br)$. Third, we embed phase-curvature into the FPM forward model as described by \eqref{eqn:O_nearfield} and \eqref{eqn:quadratic_phase}, leading to:
\begin{equation}
    O_{\mathrm{init}}(\bk) = \fft{ o_{\mathrm{init}}(\br)  Q(\br) },
\end{equation}
which serves as the initial estimate for the FPM reconstruction. Once reconstructed, the spectrum $O(\bk)$ is back-propagated into the sample plane, where the quadratic phase is conjugated out to obtain the reconstructed sample $o(\br)$:
\begin{equation}
    o(\br) = \ifft{ O(\bk) } Q^*(\br).
\end{equation}
The initialization and conjugation process is illustrated by \figref{fig:fpm_reconstruction}(b) in the context of segmentation-based reconstruction. As will be shown in the next section, this initialization procedure is a crucial ingredient for stable reconstruction of both narrow- and wide-field images in FPM.

\section{Results}
In this section, we will validate the proposed phase-curvature model and its correction by simulations and experimental reconstructions based on a single-lens, non-telecentric imaging system. Firstly, we will show that phase-curvature appears during FPM reconstructions in both simulated and experimental data. Next we will show that the issue becomes even greater when we consider compact microscopes due to short propagation distances. In addition, while phase-curvature is a phase only aberration, it will affect the overall algorithmic convergence which is undesirable even if aberrated phase reconstruction can be tolerated. Lastly, while phase-curvature may not be significant for small FoVs the reconstructions can still fail due to poor initialization, imposing a practical limit of how small the segments can be.

In summary, we will show the following:
\begin{itemize}
    \setlength\itemsep{0em}
    \item Existence of the phase-curvature in both real and simulated data.
    \item The presence of residual phase-curvatures when using segmentation-based reconstructions.
    \item phase-curvature elimination using our proposed method for any FoV size.
    \item Compromised reconstruction convergence—and in some cases failure—if our method is not used.
    \item Amplitude reconstruction degradation if phase-curvature is not corrected.
\end{itemize}

\subsection{Numerical simulations}

\begin{figure}[t!]
    \centering
    \includegraphics[width=0.8\linewidth,trim={0 0 0 0}]{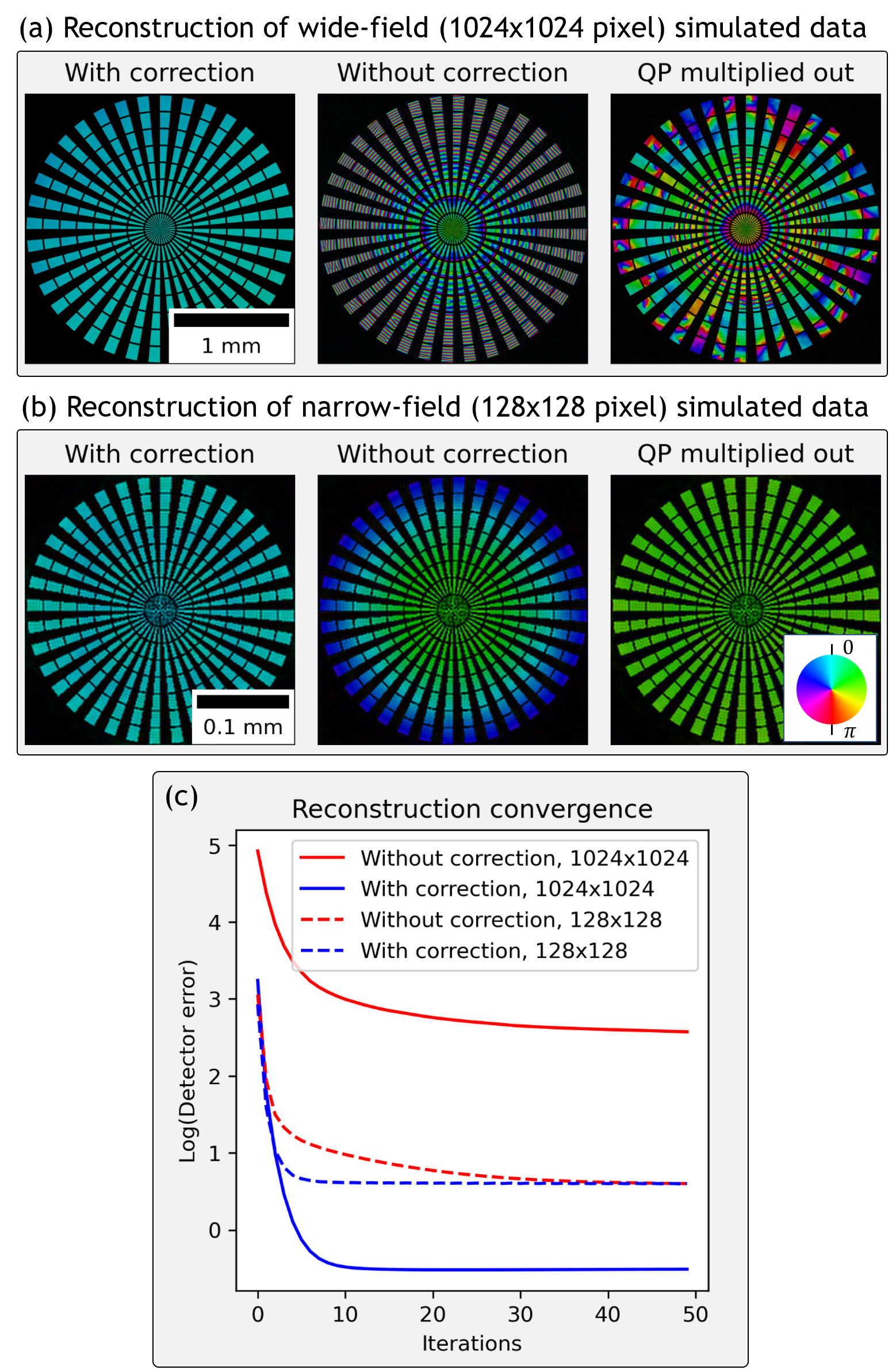}
    \caption{FPM data was simulated for FOV sizes of (a) $128\times128$ and (b) $1024\times1024$ pixels. While phase-curvature varies with FOV size, it is significant even for the small FOV. For wide-field reconstruction, phase-curvature is large enough that severe phase wrapping occurs. In both cases, the computational correction method was able to reconstruct images free of phase-curvature compared to basic FPM reconstructions without the proposed correction method. Also, even in ideal imaging conditions, the phase-curvature is difficult to reconstruct, which is why it cannot be multiplied out post reconstruction as shown in (a) necessitating our proposed initialization. We also show in (c) that without computational curvature correction, algorithmic convergence is severely impeded due to the presence of severe phase wrapping.}
    \label{fig:simulations}
\end{figure}

To validate the presence of phase-curvature and the proposed correction strategy, we carried out simulations using spherical illumination wavefronts. The wavefronts transmitted and scattered by the sample, were then propagated to the image plane using Fresnel diffraction, rather than by using a pre-defined image-formation model such as the one in \eqref{eqn:image_formation}. In doing so, the phase-curvature appears purely as a result of the wave-propagation phenomena in the Fresnel approximation. The simulations can be summarized by the following steps:
\begin{itemize}
    \setlength\itemsep{0em}
    \item Create a scattered wavefront as a result of sample illumination (\eqref{eqn:Thin sample approximation}), which will also impart a phase shift resulting in Fourier-domain spectrum translation (due to angular illumination) as described in \eqref{eqn:spherical_final2}.
    \item Propagate the wavefront from the sample to the lens plane using a Fresnel diffraction propagator from \eqref{eqn:fresnel_diffraction}.
    \item Perform frequency filtering by the pupil $P(\bk)$ and apply a phase transformation due to interaction with the lens ~\cite{Goodman2005Introduction}.
    \item Propagate to the detector plane using another Fresnel propagator from \eqref{eqn:fresnel_diffraction}, transforming the diffracted fields back into the spatial domain. 
    \item Perform incoherent image detection by $|\cdot|^2$.
\end{itemize}

We simulated imaging of a Siemens star target with a FoV size of $128\times128$ ($0.3mm^2$) pixels (\figref{fig:simulations}(a)) and $1024\times1024$ ($2.3mm^2$) pixels (\figref{fig:simulations}(b)), based on the optical design described in \secref{sec:Optical configurations} and \figref{fig:optical_design}(a). The data consisted of images recorded at $49$ illumination angles, and the expected phase-curvatures are shown in \figref{fig:optical_design}(a) for both FoV areas. The simulated Siemens star targets were amplitude-only samples, hence, no phase-curvature should be present in the reconstructions. As predicted, both narrow-field and wide-field reconstructions in \figref{fig:simulations}(a,b) suffer from illumination and wave-propagation induced curvature, which is correctly eliminated with the proposed correction method.

We can also see from \figref{fig:simulations}(b) that phase-curvature can be largely eliminated by multiplying it out post-reconstruction (without prior initialization). However, for larger FoVs, post-reconstruction removal is no longer possible as shown in \figref{fig:simulations}(a), producing residual phase artefacts. Failure to address phase-curvature prior to the reconstruction process also impedes reconstruction convergence, as shown in \figref{fig:simulations}(c). Through an appropriate initialization scheme, the initial guess is closer to the global minima, resulting in reduced computational requirements to reach the optimal solution. \figref{fig:simulations}(c) also shows that reconstruction convergence between  wide-field and narrow-field reconstructions is equivalent as long as phase-curvature correction is used, making our method a suitable substitute for segmentation-based reconstructions.

\subsection{Optical configurations}
\label{sec:Optical configurations}

\begin{figure}[t]
    \centering
    \includegraphics[width=\linewidth,trim={0 0 0 0}]{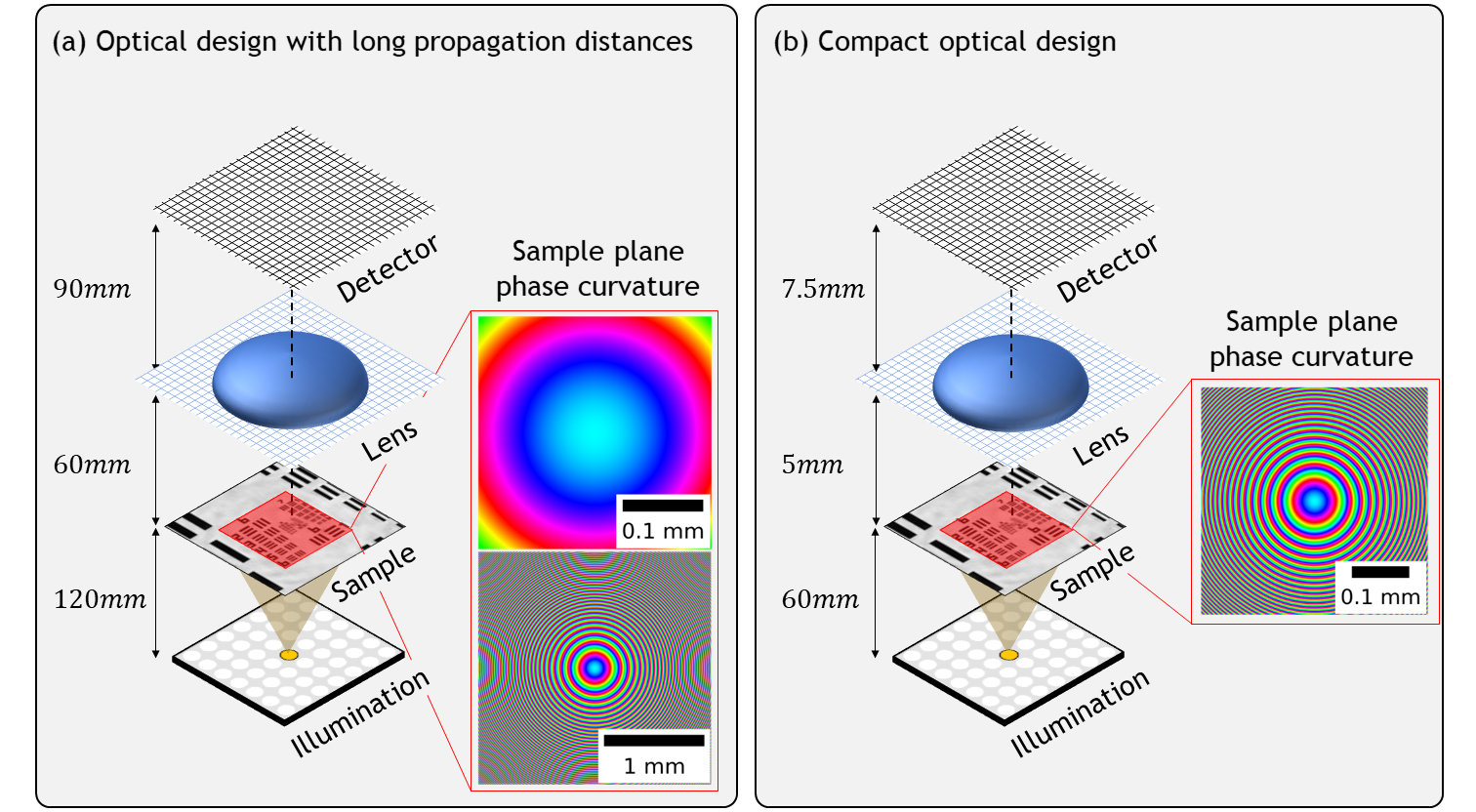}
    \caption{Data from two illustrative microscopes will be reconstructed in this manuscript. The prototype in (a) has long propagation distances, representative of typical low-NA microscopes used for FPM. The compact prototype in (b) has the same magnification of 1.5 as in (a), but here the propagation distances between the planes are about an order-of-magnitude shorter. As a result, a significantly increased phase-curvature is observed compared to (a), across a similar FoV area.}
    \label{fig:optical_design}
\end{figure}

We will validate our finding with experimental data reconstructions at various FoV sizes using two non-telecentric optical setups. 

The microscope illustrated in \figref{fig:optical_design}(a) has sample-to-lens distance,  and  LED-to-sample distance $u=60mm$  and  $z=120mm$ respectively, which are representative of a low-NA, long-working distance experimental setup. The microscope used a $8mm$ diameter and $36mm$ focal length achromatic lenses from “Edmund optics”. For image recording, we used “DMM 37ux264-ML” $2448\time2048$ 5-megapixel monochrome sensor boards with a $3.45\mu m$ pixel size from “The Imaging Source”.  The experimental data was captured with 441 illumination directions from $21 \times 21$ LEDs (“Adafruit LED array”) with a wavelength of $\lambda=630nm$ and $5mm$ pitch of the LEDs. Given these experimental parameters, the numerical aperture was $0.065$, resulting in the raw image resolution of $9.6\mu m$. By synthesizing $21 \times 21$ angularly illuminated images through FPM reconstruction, the synthetic numerical aperture was $0.365$ equivalent to $1.45\mu m$ resolution, which we demonstrated in~\cite{Aidukas2021High, Aidukas2021Next}.

We also demonstrate the proposed method on a compact, low-cost experimental setup, illustrated in \figref{fig:optical_design}(b). The optical design parameters and data are publicly available in~\cite{Aidukas2019Low}. The low-cost microscope was equipped with a Raspberry Pi V2 NOIR camera module ($8$-megapixels, $1.12\mu m$ pixel size) which also contains a $3mm$ focal-length camera lens. The frequency overlap of ~$70\%$ (required for FPM reconstruction) was obtained by placing the Unicorn HAT HD $16\times16$ LED array ($3.3mm$ pitch) $60mm$ below the sample stage. The low-resolution microscope has $0.15$ NA and a $5mm$ working distance, whereas the synthetic NA achieved after FPM reconstruction was $0.55$ (see~\cite{Aidukas2019Low}).

\subsection{Image reconstruction for a standard microscope}

\begin{figure}[t!]
    \centering
    \includegraphics[width=\linewidth,trim={3cm 0cm 3cm 0cm},clip]{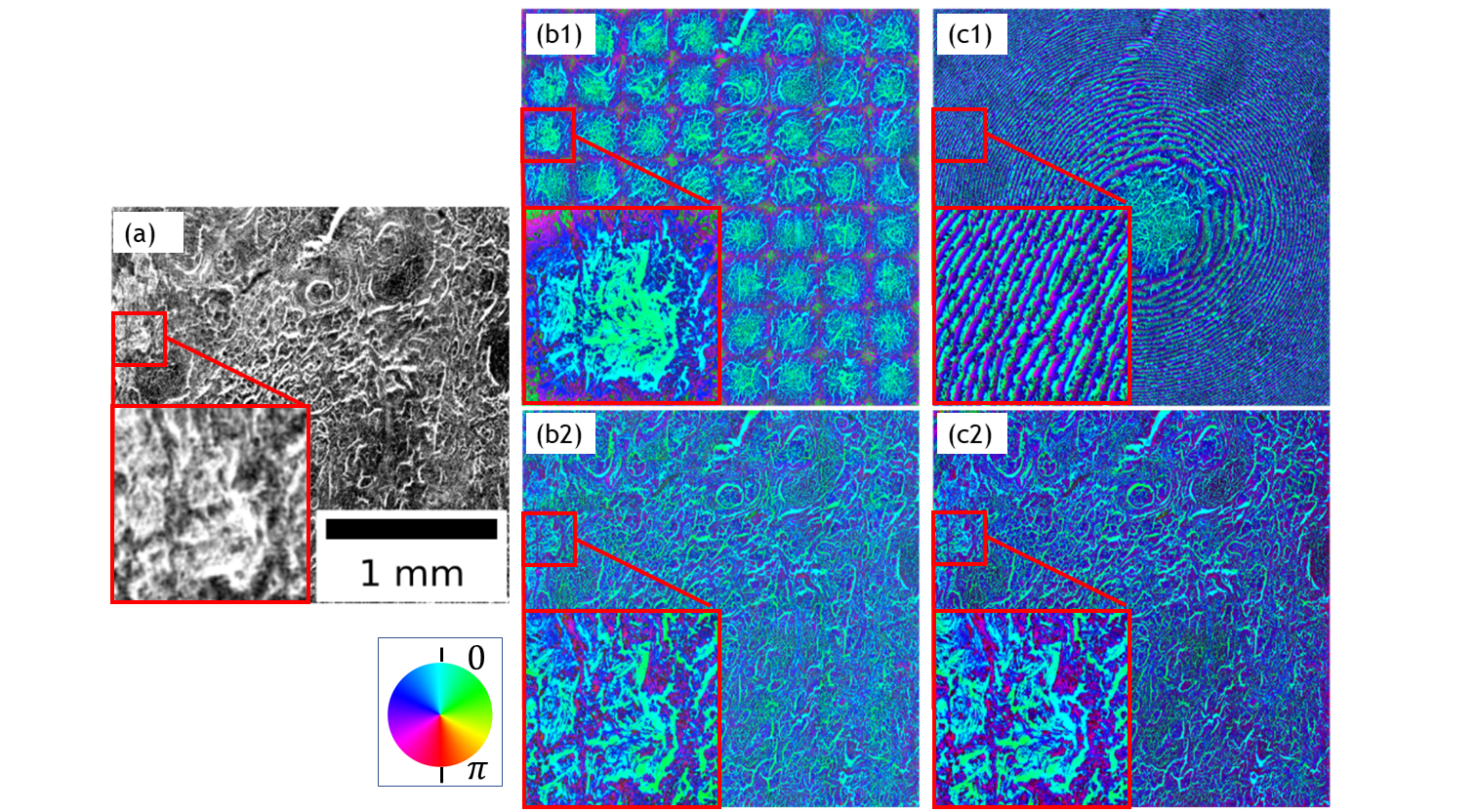}
    \caption{The raw image of a Lung Carcinoma sample $1024\times1024$ pixel FoV is shown in (a). When the FoV is divided into $128\times128$ pixel tiles, the phase-curvature is minimized, but not completely eliminated in the segmentation-based reconstruction shown in (b1), necessitating the need for correction in (b2). In segmentation-free reconstruction, phase-curvature is severely wrapped due to wide-FoV of $1024\times1024$ pixels shown in (c1) and the corresponding correction in (c2). In both narrow and wide FoV reconstructions, our proposed method is able to eliminate phase-curvature.}
    \label{fig:microscope_recons}
\end{figure}

We illustrate the need for phase-curvature correction for both segmentation-free and segmentation-based reconstructions using a lung carcinoma sample shown in \figref{fig:microscope_recons}(a). In all reconstruction comparisons with and without phase-curvature correction, equivalent reconstruction parameters were used. Also, we will refer to various segment sizes by the pixel count rather than FoV dimensions in SI units. The following FoVs will be used:
\begin{itemize}
    \setlength\itemsep{0em}
    \item $32\times32$ pixels - \SI{0.07}{\milli\metre} $\times$ \SI{0.07}{\milli\metre}.
    \item $128\times128$ pixels - \SI{0.3}{\milli\metre} $\times$ \SI{0.3}{\milli\metre}.
    \item $1024\times1024$ pixels - \SI{2.3}{\milli\metre} $\times$ \SI{2.3}{\milli\metre}.
\end{itemize}

In the segmentation-based reconstruction shown in  \figref{fig:microscope_recons}(b1-b2), the $1024\times1024$ pixels for the full FoV were divided into smaller $128\times128$ pixel segments to reduce spatially varying aberrations and phase-curvature. Once reconstructed, all segments were tiled together into a single wide-field, high-resolution image. Without correction (\figref{fig:microscope_recons}(b1)), each reconstructed tile contains minor phase-curvature, resulting in a distorted phase map of the sample. Once corrected (\figref{fig:microscope_recons}(b2)), the tiles can be seamlessly stitched together without any visible phase discontinuities.

In the segmentation-free reconstructions from \figref{fig:microscope_recons}(c1-c2), the missing phase-curvature results in significant phase aberrations based on results in \figref{fig:microscope_recons}(c1). While it would be convenient to simply use conventional FPM methods and multiply out the phase-curvature post-reconstruction, the inconsistencies in panel (c1) would remain due to visible dissimilarity from the model in \figref{fig:optical_design}(a). In \figref{fig:microscope_recons}(c2) we followed our proposed initialization approach on the large field of view. This reconstruction is free of the phase artefacts observed in panel (c1). Moreover, the reconstruction in panel (c2) is consistent with the segmentation-based reconstruction in (b2). 

\begin{figure}[t!]
    \centering
    \includegraphics[width=\linewidth,trim={0cm 0cm 0cm 0},clip]{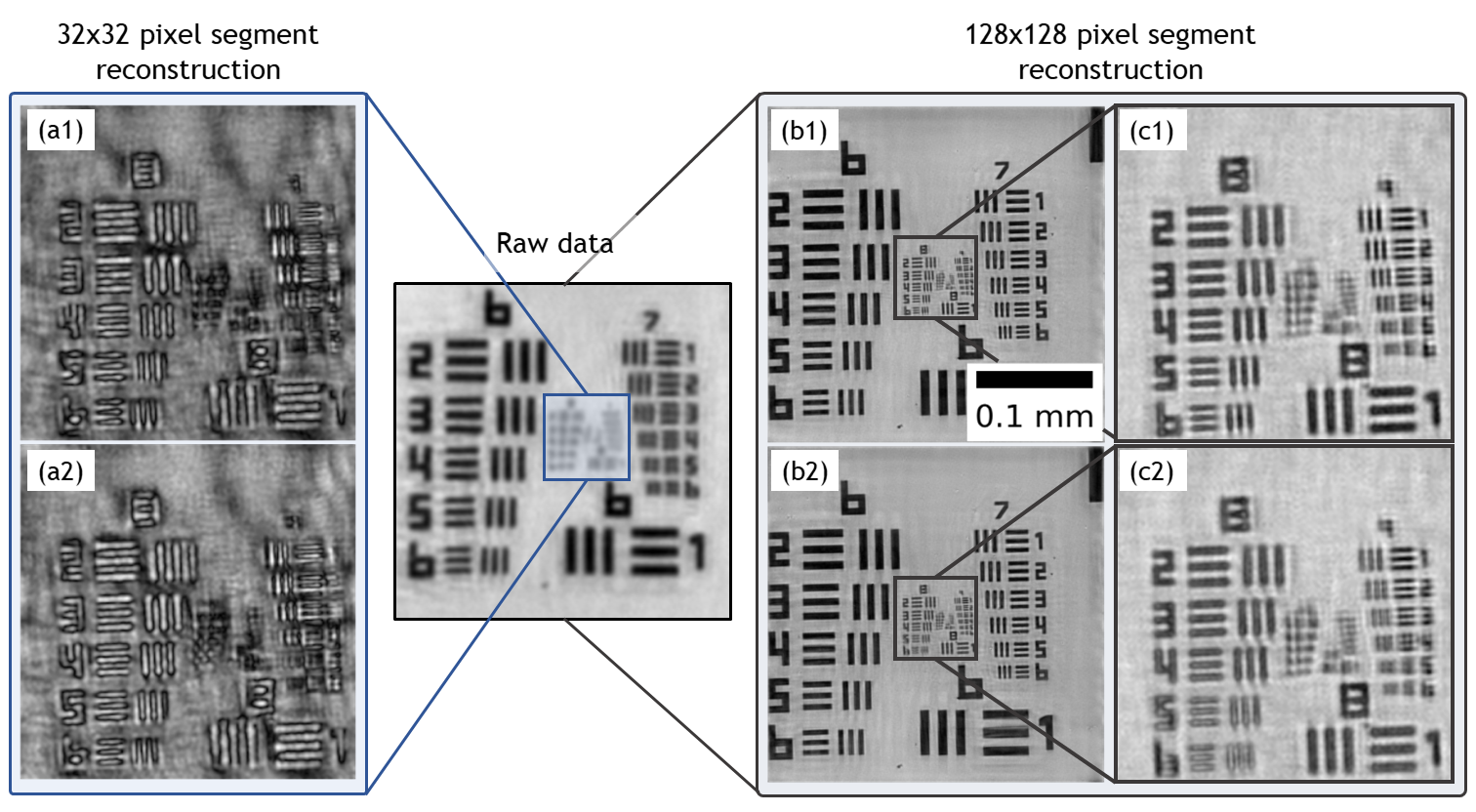}
    \caption{Amplitude reconstructions of a USAF target 32×32 pixel FoV are shown in (a1) and (a2) with and without phase-curvature correction respectively.  Once the FoV size is reduced too much, the algorithmic convergence suffers, which is why FoV size reduction is not always feasible to eliminate the phase-curvature. By reconstructing a wider 128×128 pixel FoV area with and without phase-curvature correction in (b1) and (b2) respectively, the reconstruction quality is improved, despite the presence of a more significant phase-curvature. Moreover, the zoomed in sections show that with phase-curvature correction in (c1) the bars are less blurry when compared to (c2).}
    \label{fig:usaft_recons}
\end{figure}

While it might be tempting to reduce the segment size even further to eliminate the presence of phase-curvature, it can come at a cost of degraded algorithmic convergence. We demonstrate this in \figref{fig:usaft_recons} where a USAF resolution target $32\times32$ pixel central sub-area was selected from $128\times128$ pixel FoV. Again, all reconstructions are performed with an equivalent iteration number as well as other parameters. Note that we show only the amplitude reconstructions to demonstrate that phase-curvature is not limited only to phase reconstruction. The small-FoV reconstructions in \figref{fig:usaft_recons}(a1-a2) indicate complete failure of the reconstructions, which is not the case for wider FoVs shown in \figref{fig:usaft_recons}(b1) (with correction) and \figref{fig:usaft_recons}(b2) (without correction). Also, looking at the zoomed in sections in \figref{fig:usaft_recons}(c1-c2) it is clear that amplitude reconstructions are less blurry in panel (c1) where the phase-curvature correction method was used. In general, the presence of more scatterers within a wider FoV will improve algorithmic convergence, especially for sparse samples such as cell cultures. This is why reconstruction of larger FoVs is desirable, necessitating the proposed phase-curvature correction.

\begin{figure}[t!]
    \centering
    \includegraphics[width=\linewidth,trim={3cm 0cm 3cm 0},clip]{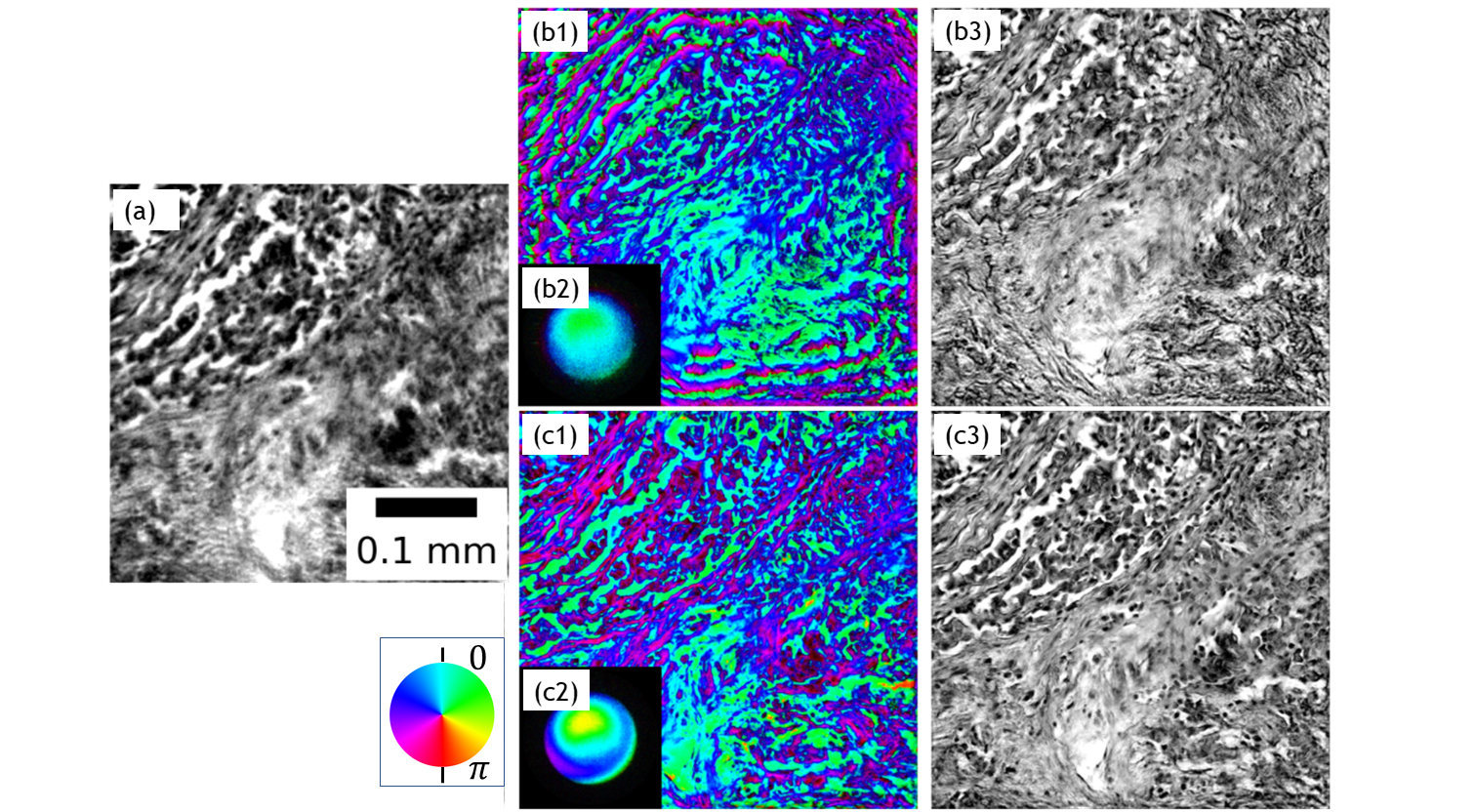}
    \caption{Reconstructions of a lung carcinoma sample in (a) using a compact, non-telecentric microscope design. While phase reconstructions without curvature correction are aberrated (b1), the reduced algorithmic convergence results in poor pupil (b2) and amplitude (b3) reconstructions. When phase-curvature is initialized prior to FPM reconstructions, the phase (c1), pupil (c2) and amplitude (c3) reconstruction quality is significantly improved.}
    \label{fig:low_cost_recons}
\end{figure}

\subsection{Image reconstruction for a compact microscope}
We now demonstrate reconstructions for a compact optical setup described in \secref{sec:Optical configurations}, using a lung carcinoma sample $768\times 768$ pixel (\SI{0.57}{\milli\metre} $\times$ \SI{0.57}{\milli\metre}) segment shown in \figref{fig:low_cost_recons}(a). The phase-curvature is expected to be more severe due to shorter propagation distances (\eqref{eqn:quadratic_phase}), which is validated by reconstructions in \figref{fig:low_cost_recons}(b1), despite having a $16\times$ smaller FoV area compared to \figref{fig:microscope_recons}(c2). We also note that the presence of phase-curvature significantly impacted not only phase reconstructions, but also pupil aberration and sample amplitude recovery, shown in \figref{fig:low_cost_recons}(b2-b3). With phase-curvature correction, the aforementioned problems are eliminated, as demonstrated by \figref{fig:low_cost_recons}(c1-c3). These results illustrate that phase-curvature correction is important not only for compact microscopes (due to significantly shorter propagation distances), but also for low-cost microscopes which suffer from poor data quality due to higher optical aberrations and lower signal-to-noise ration. In such instances, any additional inconsistencies between the forward model and recorded data can severely affect the FPM algorithmic ability to converge properly, which can be alleviated with the proposed correction.

\section{Discussion}
We have shown that the phase-curvature present in finite-conjugate microscopes appears in the reconstructed images. This is normally mitigated by reconstructing extended images using a segmentation-based approach or by using telecentric optics. A telecentic lens is designed to offer the same magnification, irrespective of longitudinal or axial distance from the lens, but requires complicated high-cost lenses. Telecentric lenses can be used not only as microscope objectives, but also to transform a spherical illumination wavefronts into plane-waves for FPM~\cite{Zhu2020Single}. However, the strength of FPM lies in its innate ability to offer high-performance imaging with extremely low-cost optical components. In such case, the use of expensive telecentric, aberration-free optics goes against the ethos of FPM. Our phase correcting algorithm provides an improved efficiency that is particularly important for achieving the potential for low-cost microscopy. Phase-curvature correction is also important to Fourier ptychography at x-ray wavelengths~\cite{Wakonig2019X}, where manufacture of even simple focusing optics is extremely challenging. The absence of telecentric optics, means that phase-curvature is unavoidable.

In some instances, it can be difficult to know every optical design parameter to accurately model the phase-curvature for correction. For example, we have assumed that LEDs are ideal point source illuminators producing spherical-wavefronts. However, LED arrays can have plastic lenses covering each LED to increase the directionality of the illumination. Similarly, microscope objectives can be close to, but not quite, telecentric, rather than satisfying one of the extremes that we presented in this paper. Fortunately, we have observed that even with incorrect initial phase-curvature estimates, the algorithm is pushed closer to the global minima, yielding improved reconstruction quality. In addition, computational optimization can be used to find the most optimal phase-curvature, similar to methods used in digital holographic microscopy~\cite{Zuo2013Phase,Ferraro2003Compensation}. Lastly, if required, phase-curvature within complicated optical designs can be estimated by using ABCD matrix based models~\cite{Bandres2009Paraxial,GuizarSicairos2006Generalized}.

Given the dependence of phase-curvature on FoV, segmentation-based reconstruction is an exceptionally useful tool for minimization of phase aberrations. While such reconstruction method was intended for non-planar illumination wavefront correction, it has proven exceptionally useful for non-telecentric optical systems. As we have shown, even with segmentation-based reconstruction, the phase-curvature can still be visible in the reconstructed images. While curvature can be mitigated by an even smaller segment FoV, the smaller the real space segment, the coarser the sampling of the pupil aberrations and reconstructed spectrum will be. As we have shown, for small enough segments, reconstruction quality can begin to deteriorate. Also, if the segment dimension is halved (along rows and columns), the total number of segments required to divide the total image FoV quadruples. The number is even larger since each segment must overlap with each other for seamless image stitching. By having a larger number of smaller arrays to process, data loading and pre-processing overhead starts to increase, whereas the parallelized graphical processing unit (GPU) computation efficiency is lost.

Lastly, if no phase wrapping is present, the phase-curvature can be eliminated during image post-processing by, e.g., computational background removal. However, such arbitrary alteration compromises the quantitative phase reconstructions. Given that our method requires only two multiplications prior and after the reconstruction, there is no drawback in using it. As we have shown, not only is the phase aberration removed, but computational convergence and amplitude reconstructions are also improved.

\section{Conclusion}
We have shown that FPM reconstruction initialization with the expected phase-curvature model and post-reconstruction removal provides aberration free quantitative phase images. Our method can accommodate the more general, and pertinent case of non-telecentric optical designs together within illumination phase-curvature into FPM forward model. Surprisingly, this rather simple but important modelling step has not been reported to date. The initialization method proposed here is simple and computationally efficient, requiring only two multiplications prior to and following reconstruction. Not  only does the correction remove the phase-curvature artefacts, but it also improves algorithmic convergence, since reconstruction of the phase-curvature itself is no longer required. The reduced computational burden is especially important when using highly aberrated, low-cost optics~\cite{Aidukas2019Low}, where the need to recover both the aberrations and phase-curvature is likely to result in reconstruction failure. Moreover, if image-formation can be assumed spatially-invariant, segmentation-free reconstruction can be performed without visible phase-curvature, which is important for quantitative phase imaging. In summary, the  proposed method improves algorithmic convergence and bypasses time-consuming stitching as well as phase synchronization of adjacent sample regions in segmentation-based FPM.

\section{Disclosures}
The authors declare no conflicts of interest.

\section{Data Availability}
We provide Jupyter notebooks for data simulation and reconstruction, together with datasets and complete Fourier ptychography reconstruction software in \url{https://doi.org/10.6084/m9.figshare.19137611}.

\bibliography{manuscript}

\clearpage

\section{Supplementary material}
The forward model in Fourier ptychography is based on the classic image-formation model found in most optical textbooks~\cite{Goodman2005Introduction}. We revisit the FPM model and address two common approximations. The first approximation assumes plane-wave illumination, but we will show that spherical-wave illumination can also be used if an additional phase-curvature term is taken into account. The second approximation neglects the quadratic phase term between the object and pupil planes, which holds if~\cite{Goodman2005Introduction}:
\begin{quote}
    \begin{enumerate}
        \item The object is located on the surface of a sphere of radius $r$ centered on the point where the optical axis pierces the thin lens.
        \item The object is illuminated by a spherical-wave that is converging towards the point where the optical axis pierces the lens.
        \item The phase of the quadratic phase factor changes by an amount that is only a small fraction of a radian within the region of the object that contributes significantly to the field at the particular image point.
    \end{enumerate}    
\end{quote}
While conditions 1 and 2 are rarely satisfied in practice, condition 3 is typically assumed to be true~\cite{Goodman2005Introduction}. However, such approximation is not always valid, as shown in the main text. We will demonstrate the presence of the quadratic phase exponential in FPM forward model when using a non-telecentric imaging system. In addition, we show that the quadratic phase exponential vanishes in telecentric systems.

\subsection{Spherical-wave illumination}\label{section:plane_waves}

\begin{figure}[h!]
    \centering
    \includegraphics[width=\linewidth,trim={0cm 0cm 0cm 0},clip]{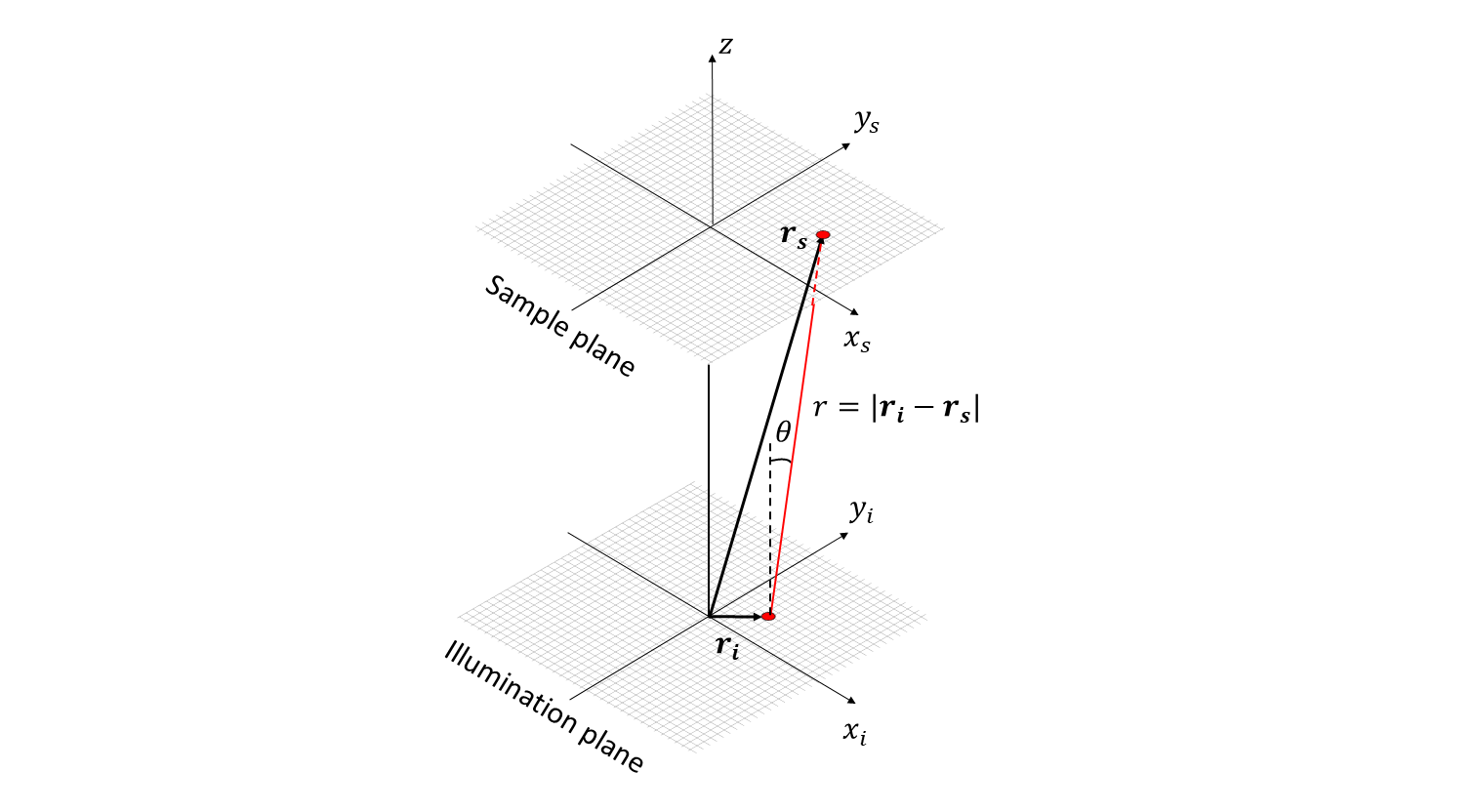}
    \caption{Wave-propagation geometry between two parallel planes.}
    \label{fig:prop_geometry}
\end{figure}

The illumination in Fourier ptychography is modelled as a tilted plane-wave defined as a “phase ramp”:
\begin{equation}
    p(\br_s) = \expp{i \bk_i \br_s} = \expp{i(k_{x,i} x_s + k_{y,i} y_s)}.
    \label{eqn:fpm_planewave}
\end{equation}
$\br_s=(x_s,y_s)$ defines the sample plane coordinates to which the plane-wave is propagated. A plane-wave illumination with a k-vector $\bk_i$ shifts  the sample spectrum by the same $\bk_i$ in the Fourier plane. Both $\bk$ and $\bk_i$ are assumed to have the same magnitude $|\bk|=|\bk_i|=k=2\pi/\lambda$. While plane-waves are idealized mathematical objects, the plane-wave assumption is a good approximation for sufficiently small FOVs in real-world applications. If the approximation  is broken, the corresponding translation of the sample spectrum can no longer be described by a simple shift in the Fourier plane. The use of non-planar illumination requires a more general description, which can be provided by assuming illumination from a point-source. In this case, each LED can be assumed as a source of spherical-waves~\cite{Goodman2005Introduction}, given by:
\begin{equation}
    p'(\br_s)  = \frac{\expp{ikr}}{r}.
    \label{eqn:spherical_full}
\end{equation}
$r$ is the displacement between the source plane $\br_i=(x_i,y_i)$ and the destination plane $\br_s=(x_s,y_s)$ given by the geometry in \figref{fig:prop_geometry}:
\begin{equation}
    \begin{split}
    r =& |\br_s - \br_i | = \sqrt{(x_s-x_i)^2 + (y_s-y_i)^2 + z^2} \\
    =& z\sqrt{1 + \frac{(x_s-x_i)^2+(y_s-y_i)^2}{z^2}},
    \end{split}
\end{equation}
where $z$ defines the distance from the LED array to the sample plane. To simplify the expression, we can decompose $r$ into multiple terms using a Taylor expansion and truncating higher order terms:
\begin{equation}
    \begin{split}
        r \approx& z + \frac{(x_s-x_i)^2+(y_s-y_i)^2}{2z} \\
        =& z + \frac{x_s^2+y_s^2}{2z} - \frac{x_s x_i + y_s y_i}{z} + \frac{x_i^2+y_i^2}{2z}.
    \end{split}
    \label{eqn:fpm_r_paraxial_spherical}
\end{equation}
Using this approximation for $r$, the spherical-wave expression in \eqref{eqn:spherical_full} can be written as:
\begin{equation}
    \begin{split}
        p'(\br_s)  =& \expp{ikz} \expp{ik\frac{x_s^2+y_s^2}{2z}} \expp{-ik\frac{x_s x_i + y_s y_i}{z}}\expp{ik\frac{x_i^2+y_i^2}{2z}}.
    \end{split}
    \label{eqn:spherical_full_taylor}
\end{equation}
The $1/r$ term present in \eqref{eqn:spherical_full} can be neglected, because it only provides a slowly varying intensity scaling of each detected image and is not relevant for the current discussion. Terms $\expp{ikz}$ and $\expp{ik\frac{x_{i}^2+y_{i}^2}{2z}}$ are constant with respect to sample coordinates and can be neglected, reducing \eqref{eqn:spherical_full_taylor} to:
\begin{equation}
    \begin{split}
        p'(\br_s)  =& \expp{\frac{ik}{2z}(x_s^2+y_s^2)} \expp{-ik\frac{x_s x_{i} + y_s y_{i}}{z}}.
    \end{split}
    \label{eqn:spherical_final}
\end{equation}
If Fourier-domain wave vectors are introduced by the following substitution:
\begin{equation}
    \begin{split}
        k_{x,i} = -k x_i / z, \quad k_{y,i} = -k y_i / z,
    \end{split}
    \label{eqn:fpm_k_spherical}
\end{equation}
the \eqref{eqn:spherical_final} can be written into the following form:
\begin{equation}
    \begin{split}
        p'(\br_s)  &=  \expp{\frac{ik}{2z}(x_s^2+y_s^2)}  \expp{i ( k_{x,i} x_s + k_{y,i} y_s}.
    \end{split}
    \label{eqn:spherical_final2}
\end{equation}
Comparing the plane-wave expression from \eqref{eqn:fpm_planewave} to the paraxial approximation in \eqref{eqn:spherical_final2}, the only difference is an additional quadratic phase factor:
\begin{equation}
    Q_{\text{ill}}(x_s,y_s) = \expp{\frac{ik}{2z}(x_s^2+y_s^2)}.
\end{equation}
This quadratic phase factor can be eliminated during FPM reconstruction, allowing the use of spherical-wave illumination instead of the plane-wave approximation conventionally used in FPM.

\begin{figure}[t!]
    \centering
    \includegraphics[width=\linewidth,trim={5cm 0cm 5cm 0},clip]{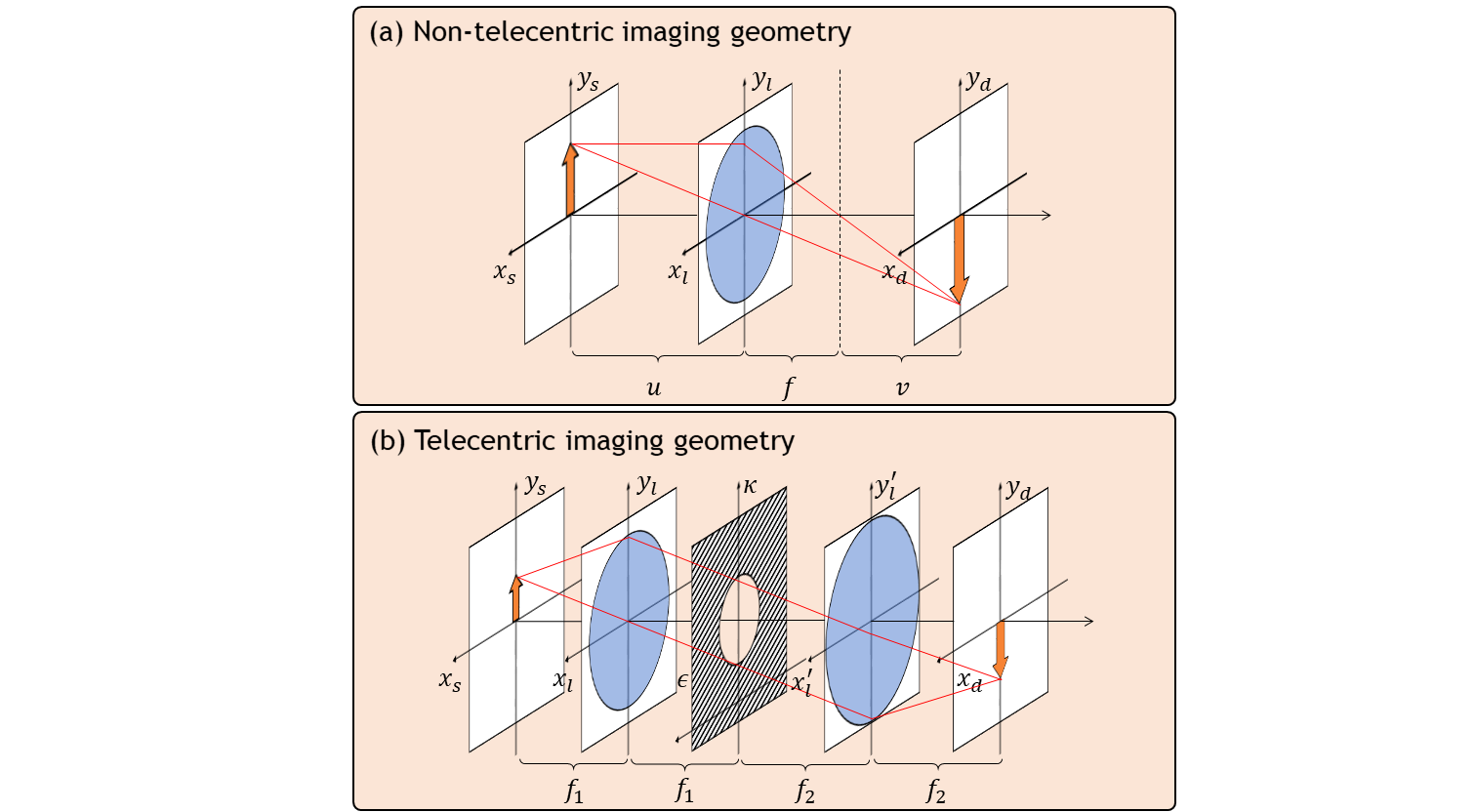}
    \caption{(a) Non-telecentric imaging geometry based on a single-lens microscope configuration. (b) Telecentric imaging geometry based on a 4f microscope configuration.}
    \label{fig:supp_optical_fig}
\end{figure}

\subsection{Image-formation model} \label{section:image-formation model}
To derive the image-formation model required for FPM, we begin by assuming the optical geometry from \figref{fig:supp_optical_fig}(a) and that the sample $o(x_s,y_s)$ is thin~\cite{Goodman2005Introduction}. The sample illuminated by spherical-waves originating from $\br_i=(x_{i},y_{i})$ with respect to the sample plane $\br_s=(x_s,y_s)$ is expressed by:
\begin{equation}
    o'(x_s,y_s)  =  \expp{\frac{ik}{2z}(x_s^2+y_s^2)} \expp{\frac{ik}{z}(x_s x_{i} + y_s y_{i})} o(x_s,y_s).
    \label{eqn:Thin sample approximation}
\end{equation}
To propagate the diffracted wavefront $o'(x_s,y_s)$ from one plane to the other, we can use a Fresnel diffraction integral~\cite{Goodman2005Introduction}:
\begin{equation}
    \begin{split}
        \psi&(x_1,y_1) =  \frac{\expp{ikz}}{i\lambda z} \expp{\frac{ik}{2z} (x_1^2+y_1^2)} 
        \iint \psi(x_0,y_0) \\& \expp{\frac{ik}{2z} (x_0^2+y_0^2)}  \expp{-\frac{ik}{u}(x_0 x_1+y_0 y_1)}\text{d}x_0\text{d}y_0,
    \end{split}
    \label{eqn:fresnel_diffraction}
\end{equation}
which describes propagation of an arbitrary wave $\psi(x_0,y_0)$ from the plane $(x_0,y_0)$ to $(x_1,y_1)$ a distance $z$ away.

By using \eqref{eqn:fresnel_diffraction}, we can propagate the diffracted wavefront $o'(x_s,y_s)$ from the sample plane $(x_s,y_s)$ to the lens plane $(x_l,y_l)$ a distance $u$ away, followed by another propagation to the detector plane $(x_d,y_d)$ a distance $v$ away. The wavefront at the detector plane can be expressed as~\cite{Goodman2005Introduction}:
\begin{equation}
    \begin{split}
        \psi(x_d,y_d) =& -\frac{\expp{ik(v+u)}}{\lambda^2 v u } \expp{\frac{ik}{2v} (x_d^2+y_d^2)}  
        \\
        & \iint \Bigg[ \iint o'(x_s,y_s) \expp{\frac{ik}{2u} (x_s^2+y_s^2)} \expp{-\frac{ik}{u}(x_l x_s+y_l y_s)}\text{d}x_s\text{d}y_s \Bigg]\\
        & P\left(\frac{kx_{l}}{u},\frac{ky_{l}}{u}\right) \expp{-\frac{ik}{v}(x_d x_l+y_d y_l)} \text{d}x_l\text{d}y_l
    \end{split}
    \label{eqn:o_to_img_single}
\end{equation}

Next, we discuss the telecentric 4f imaging configuration shown in \figref{fig:supp_optical_fig}(b), which is able to eliminate the quadratic phase terms present in a single-lens system. To obtain the image-formation model for a 4f system, we can apply \eqref{eqn:o_to_img_single} twice: once to propagate the field from the sample plane $(x_s,y_s)$ to the intermediate plane $(\epsilon,\kappa)$ and a second time to propagate towards the detector plane $(x_d,y_d)$. Such configuration assumes that the first and second lenses have the focal lengths $f_1$ and $f_2$ respectively and that there is a limiting aperture in the intermediate plane $(\epsilon,\kappa)$ defined by the pupil function $P\left(k\epsilon/f_1,k\kappa/f_1\right)$. The wavefront in the detector plane can be written as~\cite{Mertz2019Introduction}:
\begin{equation}
    \begin{split}
        \psi_t(x_d,y_d) =&  \frac{\expp{2ik(f_1+f_2)}}{\lambda^2 f_1 f_2 } \\
        &\iint \Bigg[\iint o'(x_s,y_s)  \expp{-\frac{ik}{f_1}(\epsilon x_s+ \kappa y_s)}  \text{d}x_s\text{d}y_s\Bigg]\\ 
        & P\left(\frac{k\epsilon}{f_1},\frac{k\kappa}{f_1}\right) \expp{-\frac{ik}{f_2}(\epsilon x_d+ \kappa y_d)}  \text{d}\epsilon \text{d}\kappa.
    \end{split}
    \label{eqn:o_to_img_4f}
\end{equation}
The subscript $t$ in $\psi_t(x_d,y_d)$ refers to “telecentric” imaging. Compared to \eqref{eqn:o_to_img_4f}, we see the presence of additional quadratic phase exponentials in \eqref{eqn:o_to_img_single}. This is why the reconstructed phase maps in FPM suffer from phase aberrations when using non-telecentric imaging systems.

Both \eqref{eqn:o_to_img_single} and \eqref{eqn:o_to_img_4f} can be further simplified by writing the integrals inside the square brackets as Fourier transformations. For  \eqref{eqn:o_to_img_single} we get:
\begin{equation}
    \begin{split}
        & \iint o'(x_s,y_s) \expp{\frac{ik}{2u} (x_s^2+y_s^2)} \expp{-\frac{ik}{u}(x_l x_s+y_l y_s)}\text{d}x_s\text{d}y_s =\\
        &=\iint o(x_s,y_s) \expp{\frac{ik}{2z}(x_s^2+y_s^2)}  \expp{\frac{ik}{2u} (x_s^2+y_s^2)} \\
        & \expp{-i\left(\left(\frac{kx_{l}}{u}-\frac{kx_{i}}{z}\right)x_s + \left(\frac{ky_{l}}{u}-\frac{ky_{i}}{z}\right)y_s\right)} \text{d}x_s\text{d}y_s \\
        &= \fft{ o(x_s,y_s) \expp{\frac{ik}{2z}(x_s^2+y_s^2)}  \expp{\frac{ik}{2u} (x_s^2+y_s^2)}}\\
        &= O\left(\frac{kx_{l}}{u}-\frac{kx_{i}}{z}, \frac{ky_{l}}{u}-\frac{ky_{i}}{z}\right),
    \end{split}
    \label{eqn:fpm_O}
\end{equation}
and similarly for \eqref{eqn:o_to_img_4f}:
\begin{equation}
    \begin{split}
        & \iint o'(x_s,y_s) \expp{-\frac{ik}{f_1}(\epsilon x_s+ \kappa y_s)}\text{d}x_s\text{d}y_s =\\
        &= \fft{ o(x_s,y_s) \expp{\frac{ik}{2z}(x_s^2+y_s^2)}}\\
        &= O\left(\frac{k\epsilon}{f_1}-\frac{kx_{i}}{z}, \frac{k\kappa}{f_1}-\frac{ky_{i}}{z}\right).
    \end{split}
    \label{eqn:fpm_O_4f}
\end{equation}
Here $O$ is the diffracted sample spectrum, translated by the angular illumination. Note that in the case of telecentric imaging, the quadratic exponential proportional to $u$ is no longer present.

Lastly, the integrals outside the square brackets in \eqref{eqn:o_to_img_single} and \eqref{eqn:o_to_img_4f} can be written as another Fourier transformation in the same manner as was done in \eqref{eqn:fpm_O}. By introducing the magnification $M=v/u$, $\bk = (kx_l/u, ky_l/u)$ and $\bk_i = (kx_i/z, ky_i/z)$ the wavefront for a non-telecentric imaging system from \eqref{eqn:o_to_img_single} can be written as:
\begin{equation}
    \begin{split}
         \psi(x_d,y_d)=& -\frac{\expp{ik(v+u)}}{4\pi^2 M} \expp{\frac{ik}{2v} (x_d^2+y_d^2)} \fft{P\left(\bk \right) O\left(\bk-\bk_i\right) }.
    \label{eqn:img_plane_nontele}
    \end{split}
\end{equation}
Similarly, the wavefront of a telecentric imaging system from \eqref{eqn:o_to_img_4f} can be simplified by introducing $M=f_2/f_1$, $\bk = (k \epsilon/f_1, k \kappa/f_1)$ and $\bk_i = (kx_i/z, ky_i/z)$:
\begin{equation}
    \begin{split}
        \psi_t(x_d,y_d) =& -\frac{\expp{2ik(f_1+f_2)}}{4\pi^2M}\fft{P\left(\bk \right) O\left(\bk-\bk_i\right) }.
    \label{eqn:img_plane_tele}
    \end{split}
\end{equation}
Ignoring the constant scaling factor $1/M$, we obtain the typical FPM forward model as the intensity of the wavefront in the detector plane $\left| \psi(\br) \right|^2$
\begin{equation}
    I_i(\br)  =  \left| \fft{ P(\bk) O(\bk-\bk_i)} \right|^2.
    \label{eqn:fpm_forward_I}
\end{equation}
\eqref{eqn:fpm_forward_I} applies to both telecentric and non-telecentric imaging system. What does differ, however, is the expression of the sample spectrum based on \eqref{eqn:fpm_O} and \eqref{eqn:fpm_O_4f}:
\begin{align*}
    O(\bk) &= \fft{ o(x_s,y_s) \expp{\frac{ik}{2z}(x_s^2+y_s^2)} } &\text{(telecentric)}\\
    O(\bk) &= \fft{ o(x_s,y_s) \expp{\frac{ik}{2z}(x_s^2+y_s^2)}  \expp{\frac{ik}{2u} (x_s^2+y_s^2)}} &\text{(non-telecentric)}.
\end{align*}
In both cases we have a phase-curvature due to spherical-wavefronts:
\begin{equation}
    Q_{\text{ill}}(x_s,y_s) = \expp{\frac{ik}{2z}(x_s^2+y_s^2)}.
\end{equation}
In addition, the non-telecentric imaging system contains an additional phase-curvature term due to wave-propagation itself:
\begin{equation}
    Q_u(x_s,y_s) = \expp{\frac{ik}{2u}(x_s^2+y_s^2)}.
\end{equation}
In single-lens systems the working distance $u$ will almost inevitably be much shorter than the distance from the illumination source to the sample $z$, making $Q_u(x_s,y_s)$ the most significant source of curvature. By using a suitable microscope objective, $Q_u(x_s,y_s)$ will be eliminated (or significantly minimized), leaving only the illumination phase-curvature $Q_{\text{ill}}(x_s,y_s)$. It should be noted, that the use of condenser lenses or other collimating optics between the sample and the illumination source can remove the illumination curvature as well.

\end{document}